\documentclass[preprint,3p]{elsarticle}
\usepackage{amsmath,amssymb,braket,slashed,mathrsfs,algorithm,algpseudocode}

\journal{the arXiv}

\begin{document}

\begin{frontmatter}

    \title{QCD thermodynamics with 
    continuum extrapolated dynamical overlap fermions}
    \author[wup]{Sz.~Borsanyi}
    \author[wup,fzj,buda]{Z.~Fodor}
    \author[buda,budb]{S.D.~Katz}
    \author[wup,fzj]{S.~Krieg}
    \author[wup,fzj]{T.~Lippert}
    \author[buda,budb]{D.~Nogradi}
    \author[buda,budb]{F.~Pittler}
    \author[wup,fzj]{K.K.~Szabo}
    \author[wup]{B.C.~Toth}
\address[wup]{Department of Physics, Wuppertal
    University, Gaussstrasse 20, D-42119 Wuppertal, Germany}

\address[fzj]{IAS/JSC, Forschungszentrum J\"ulich, D-52425
    J\"ulich, Germany}

\address[buda]{Institute for Theoretical Physics, E\"otv\"os
    University, P\'azm\'any Peter s\'etany 1/A, H-1117 Budapest, Hungary}

\address[budb]{MTA-ELTE Lend\"ulet Lattice Gauge Theory Research Group, Budapest, Hungary}

\begin{abstract}

We study the finite temperature transition in QCD with 
two flavors of dynamical fermions at a pseudoscalar pion mass of 
about 350~MeV. We use lattices with temporal extent of $N_t$=8, 10 
and 12. For the first time in the literature a continuum limit is 
carried out for several observables with dynamical overlap 
fermions. These findings are compared with results obtained within 
the staggered fermion formalism at the same pion masses and 
extrapolated to the continuum limit. The presented results correspond
to fixed topology and its effect is studied in the staggered case.
Nice agreement is found between the overlap and staggered results.

\end{abstract}


\end{frontmatter}

\section{Introduction}

The QCD Lagrangian possesses an approximate 
SU(2)$_V\times$SU(2)$_A$$\times$U(1)$_V$$\times$U(1)$_A$ 
global symmetry. Both SU(2) symmetries and U(1)$_A$ are explicitly broken by
small mass effects. SU(2)$_A$ is spontaneously broken even in the massless
limit and the U(1)$_A$ is always broken on the quantum level by 
the chiral anomaly.
According to the most popular picture the QCD 
transition at non-vanishing temperatures is related to the 
restoration of the SU(2)$_A$ chiral symmetry. Though for physical quark masses the 
transition turns out to be an analytic cross-over 
\cite{Aoki:2006we} the physics of the transition is still 
determined by the remnants of the above mentioned symmetry breaking 
and its restoration.

The staggered fermion formulation is the cheapest one among all 
lattice fermion formulations used for QCD. In addition, staggered 
fermions possess a chiral symmetry even at non-vanishing lattice 
spacings. Thus, they show the most important physical feature of 
the finite temperature QCD transition already at non-zero lattice 
spacings. These two features, low CPU demand and symmetry, explain
why almost all large-scale lattice thermodynamics projects use 
staggered fermions. For many bulk quantities 
reliable quantitative results exist. These are obtained by 
controlled continuum extrapolations. E.g. the scale of the 
temperature was calculated 
\cite{Aoki:2006br,Aoki:2009sc,Borsanyi:2010bp,Bazavov:2011nk} in 
physical units and the equation of state was determined 
\cite{Borsanyi:2010cj,Borsanyi:2013bia,Bazavov:2014pvz}.

Though staggered fermions possess a chiral symmetry, it is not the 
same as that of the continuum QCD theory. In addition, there is a 
subtle procedure how this chiral symmetry is broken, how it is 
restored and how to look at it for less than four flavors by rooting.
Besides the theoretical difficulties related to staggered fermions, there 
are technical difficulties, too.  First of all, the observables, 
that are related to pion physics suffer from large discretization 
effects related to the taste symmetry violation (see e.g. 
\cite{Bellwied:2015lba} for examples). Secondly, measuring the thermal 
correlations of quarks and hadrons is notoriously difficult, due to 
the complicated valence structure of staggered quarks.

Wilson fermions do not possess any chiral symmetry at non-vanishing lattice spacings, the symmetry is 
restored only in the continuum limit. As a consequence Wilson 
fermion based thermodynamics has large cutoff effects and small 
lattice spacings are needed to carry out controlled continuum 
extrapolations~\cite{Borsanyi:2012uq,Borsanyi:2015waa}. 

Chiral lattice fermions (overlap or domain-wall) 
are ideal candidates for lattice QCD thermodynamics.
Obviously, they are much more expensive than the staggered 
discretization. Since the first dynamical overlap study~\cite{Fodor:2003bh}
there have been a number of finite temperature results using 
overlap~\cite{Cossu:2013uua,Borsanyi:2012xf}
or domain wall~\cite{Buchoff:2013nra,Chiu:2013wwa,Bhattacharya:2014ara} fermions.
These works used $N_t=6$ and/or $N_t=8$ lattice extents. Until now, no larger 
temporal extents were used and no continuum extrapolation was performed. 
The main goal of this paper is to present the very first investigation of this kind.

In this paper we continue our thermodynamic investigations with 
overlap fermions started in 2012 in Ref.~\cite{Borsanyi:2012xf}. We had two lattice spacings back 
then, $N_t$=6 and 8. Here we go far beyond that level and extend 
that work with two finer sets of lattice spacings, $N_t$=10 and 12. 
Since the action, the observables and many of the methods are the 
same as in our first paper, here we only describe them briefly and 
focus on the novel features and improvements.  Then we present 
results with three temporal extents: $N_t$=8,10 and 12 for four
observables. We carry out a continuum extrapolation based on these 
lattice spacings and compare them to staggered calculations.

The structure of this letter is the following. In sections~\ref{se:ov} and~\ref{se:st}
we briefly summarize the overlap and staggered simulation details, respectively. We present
the results in section~\ref{se:res} and then conclude. Some important algorithmic
details are discussed in the appendix.

\section{Overlap simulation details}
\label{se:ov}
Since we extend our preliminary study~\cite{Borsanyi:2012xf} we employ the same
lattice action, which is described in detail there. For completeness we
summarize our setup here:

\begin{itemize}

    \item tree level Symanzik improved gauge action with coupling parameter $\beta$

    \item two flavors of overlap quarks with the overlap operator defined as
	\begin{equation*}
	    D =\left(m_0-\frac{m}{2}\right)\left[ 1+\gamma_5 {\rm sgn}\left(\gamma_5 W(-m_0)\right)\right]+m,
	\end{equation*}
	where $m$ is the mass of the quark, $W(-m_0)$ is a two step HEX smeared
	Wilson operator with a negative mass of $-m_0=-1.3$.
	The HEX smearing parameters are $\alpha_1=0.72$, $\alpha_2=0.60$ and $\alpha_3=0.44$. 

    \item two flavors of Wilson fermions with mass $-m_0$, which are irrelevant in the continuum limit~\cite{Fukaya:2006vs}.

    \item two boson fields with mass $m_B=0.54$ and the action
	\begin{equation*}
	    \phi^\dagger \left[ W(-m_0) + i m_B \gamma_5 \tau_3 \right] \phi,
	\end{equation*}
	which term is also irrelevant in the continuum limit.

    \item ensembles are generated using the Hybrid Monte Carlo (HMC) algorithm
    with the Zolotarev approximation of the sign function~\cite{vandenEshof:2002ms,Fodor:2003bh}. HEX smearing
    is included in the HMC as described in~\cite{Durr:2010aw}.

\end{itemize}

The effect of the irrelevant fields is to disable topological sector changes
along the HMC trajectories.  Note that the action itself
does not constrain the topology, it only differs form the QCD action by
irrelevant terms.  These irrelevant terms give a delta function like
contribution to the action at the topological sector boundaries which the
continuous HMC trajectories cannot cross. Our aim is to keep the system
in the zero topological charge sector.

\begin{figure}[h]
    \centering
    \includegraphics[width=12cm]{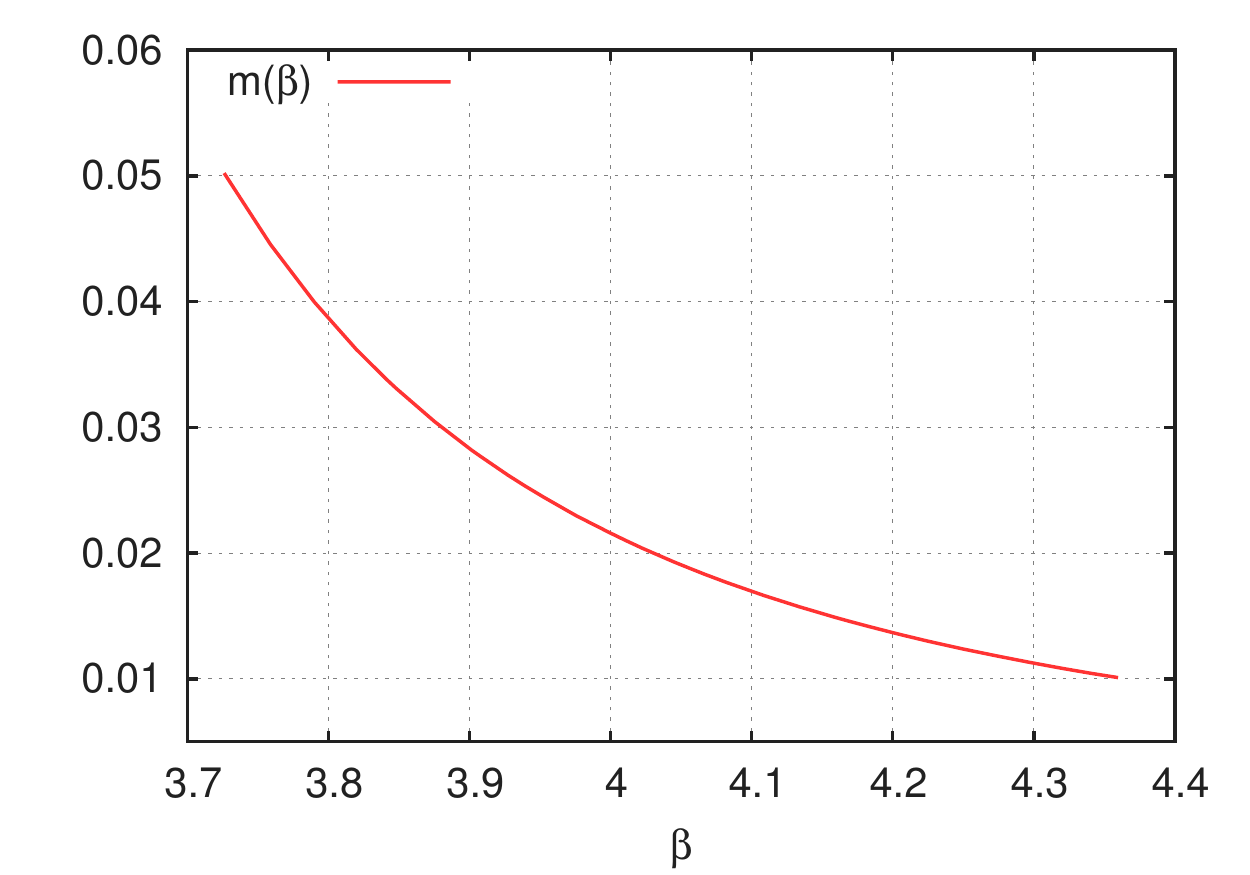}
    \caption{\label{fi:lcp} The line of constant physics used in this work.}
\end{figure}

The scale is set by the $w_0$ parameter~\cite{Borsanyi:2012zs} and our line of constant physics (LCP) is defined by the condition $m_\pi
w_0=0.312$.  In Ref.~\cite{Borsanyi:2012xf} we determined the LCP in the range
$\beta=3.6 \dots 4.1$. In this work we needed to determine the line of constant
physics (LCP) for larger $\beta$ values.  Using two additional simulations on
$32^4$ and  $32^3\cdot 48$ lattices at $\beta=4.2$ and at $4.3$, respectively, 
we obtained the LCP, which is shown on Figure
\ref{fi:lcp}.

For the renormalization of the finite temperature results we performed a series
of runs on symmetric lattices $N_s=N_t$. These are collected in Table
\ref{ta:zerot}. The $w_0$ values, that are used to convert the results to
physical units, are measured on these lattices. For the conversion we use the
value $w_0=0.1755$ fm.\footnote{Note that this choice is ambiguous since
this value corresponds to QCD with physical quark masses. Using a 
different observable to set the scale might lead to 
somewhat different results in physical units.} To check the quality of our LCP determination we also
measured the pion mass on these lattices. The $m_\pi w_0$ values are also given
in Table \ref{ta:zerot}.

\begin{table}[p]
    \centering
    \begin{tabular}{|c|c|c|c|c|}%c|}
	\hline
	$\beta$ & $N_s$ & \#traj & a[fm] & $m_\pi w_0$ \\ %& id\\
	\hline
	3.72601 & 24 &  500 & 0.207 & 0.310(1)\\% & zerot/lo/m00   \\
	3.78989 & 24 &  600 & 0.182 & 0.314(1)\\% & zerot/lo/m01   \\
	3.84097 & 24 &  800 & 0.165 & 0.312(2)\\% & zerot/lo/m02   \\
	3.95191 & 24 & 1100 & 0.135 & 0.310(2)\\% & zerot/lo/m03   \\
	4.04464 & 24 & 1700 & 0.115 & 0.305(2)\\% & zerot/hi/m00   \\
	4.13912 & 24 & 2100 & 0.096 & 0.316(5)\\% & zerot/hi/m01   \\
	4.24969 & 24 & 2300 & 0.079 & 0.330(8)\\% & zerot/hi/m02   \\
	4.35979 & 24 & 2200 & 0.066 & 0.325(17)\\%& zerot/hi/m03   \\
	4.20588 & 32 & 1000 & 0.087 & 0.299(5)\\% & zerot/hi32/m00 \\
	4.31965 & 32 & 1200 & 0.070 & 0.331(9)\\% & zerot/hi32/m01 \\
	\hline
    \end{tabular}
    \caption{\label{ta:zerot} Summary of zero temperature runs.}
\end{table}

The finite temperature lattices are simulated at three temporal extents
$N_t=8,10$ and $12$ and with aspect ratio $r=N_s/N_t=2$. Details can be found in
Table \ref{ta:fint}. The trajectories were generated in two streams. We
monitored the topological charge ($Q$) during the simulations: no sector change was
observed even though the finite stepsize integration does not necessarily forbid the change.
We measured the quark number
susceptibility and the chiral condensate on every fifth trajectory.

For the renormalization~\cite{Aoki:2006br} of the Polyakov loop, $L$,  
we carried out simulations at
$T=208$ MeV with temporal extents $N_t=4,5,6,7,8,10,12$ and $14$. We measured
the Polyakov loop to obtain the renormalization factor $F_0(\beta)=1/N_t \log L$.

\begin{table}[p]
    \centering
    \begin{tabular}{|c|c|c|}%c|}
	\hline
	$\beta$ & $N_t\times N_s$ & \#traj \\%& id \\
	\hline
	3.72601 & $8\times 16$ &  6300\\% & nt8/str01/m00\\
	3.78989 & $8\times 16$ &  7600\\% & nt8/str01/m01\\
	3.84797 & $8\times 16$ &  9000\\% & nt8/str01/m02\\
	3.90165 & $8\times 16$ & 10500\\% & nt8/str01/m03\\
	3.95191 & $8\times 16$ & 12100\\% & nt8/str01/m04\\
	3.99943 & $8\times 16$ & 13400\\% & nt8/str01/m05\\
	4.04464 & $8\times 16$ & 14400\\% & nt8/str01/m06\\
	4.10869 & $8\times 16$ & 15700\\% & nt8/str01/m07\\
	\hline
	3.84097 & $10\times 20$ &  6700\\% & nt10/str01/m00\\
	3.90810 & $10\times 20$ &  8200\\% & nt10/str01/m01\\
	3.97002 & $10\times 20$ &  9800\\% & nt10/str01/m02\\
	4.02793 & $10\times 20$ & 11200\\% & nt10/str01/m03\\
	4.08253 & $10\times 20$ & 12300\\% & nt10/str01/m04\\
	4.13412 & $10\times 20$ & 13200\\% & nt10/str01/m05\\
	4.18274 & $10\times 20$ & 13800\\% & nt10/str01/m06\\
	4.24969 & $10\times 20$ & 14100\\% & nt10/str01/m07\\
	\hline
	3.93963 & $12\times 24$ &  5800\\% & nt12/str01\_q0/m00\\
	4.01093 & $12\times 24$ &  6800\\% & nt12/str01\_q0/m01\\
	4.07720 & $12\times 24$ &  6100\\% & nt12/str01\_q0/m02\\
	4.13912 & $12\times 24$ & 10100\\% & nt12/str01\_q0/m03\\
	4.19672 & $12\times 24$ & 11000\\% & nt12/str0/m04\\
	4.24969 & $12\times 24$ & 11400\\% & nt12/str0/m05\\
	4.29763 & $12\times 24$ & 11600\\% & nt12/str0/m06\\
	4.35979 & $12\times 24$ & 10300\\% & nt12/str0/m07\\
	\hline
    \end{tabular}
    \caption{\label{ta:fint} Summary of finite temperature runs.}
\end{table}

\section{Staggered simulation details}
\label{se:st}
In order to compare the continuum results obtained from 
different discretizations we also performed a series of staggered runs. We used the
same tree level Symanzik improved gauge action with $\beta$ coupling as in the
overlap case. The fermion action is a two flavor (rooted) staggered fermion action
with four steps of stout smearing with smearing parameter $\rho=0.125$. We tuned
the quark mass to the same LCP as in the overlap case ($m_\pi w_0$=0.312). 
To this end we generated 1000-7000 configurations separated by 5 HMC trajectories
on $32^3\cdot 64$, $40^3\cdot 64$  and $48^3\cdot 64$
lattices in the range $\beta= 3.75 \dots 4.4$. These $T=0$ runs are also used for 
renormalization of the chiral observables. At non-vanishing temperature
we used $N_t$=8, 10, 12 and 16 lattices with the same fixed $r=2$ aspect ratio
as in the overlap case. This guarantees that the same physical volumes are 
compared. We chose the couplings such that they correspond to the same fixed
temperatures at each $N_t$. Each ensemble consists of 1000-2000 configurations
separated by 10 HMC trajectories.
Since in the staggered case the topology was not constrained, all our
runs sample multiple topological sectors. In the infinite volume limit 
our observables are expected to be independent of the global topological charge
but at the volumes studied in this work there may still be significant $1/V$
volume corrections~\cite{Aoki:2007ka}. To investigate these effects we also selected gauge 
configurations with zero topological charge. This makes a direct comparison
with the overlap results possible and the difference compared to having all
sectors gives an estimate of this volume dependence for each observable.
For the selection we applied a gradient flow on the gauge configurations. 
After measuring the topological charge distribution on these gauge fields 
we were always able to identify a peak in the histogram which belonged to the
$Q=0$ sector. 
We checked that this selection, and, in particular, the resulting
expectation value of the chiral condensate, was independent of the $t$
flow time if $T^2t \ge 0.0625$. For all lattice spacings and 
temperatures we used the selection rule $-0.5 < Q(t=0.0625/T^2) < 0.5$.
Note that by dropping the $Q\ne0$ sectors we lost 85\% of our statistics
below $T_c$, in the deconfined phase this loss was only about 10\%.

\section{Results}
\label{se:res}
\subsection{Continuum extrapolation}
We want to compare continuum extrapolated results of temperature
dependent observables using two discretizations. This requires a non-trivial 
analysis, since both an interpolation in $T$ and a continuum extrapolation is
needed. In case of the overlap ensembles for 
all observables we used two independent 
analyses following different strategies. These are similar to those applied in
~\cite{Borsanyi:2012uq,Borsanyi:2015waa}. In the first approach we
first interpolate the data in temperature for each lattice spacing and then 
perform a continuum extrapolation at fixed temperatures. The 
interpolation is done using a cubic spline. The continuum extrapolation is performed
as a linear fit in $1/N_t^2$ using our three lattice spacings. The continuum
extrapolations for all observables had good fit qualities.

In the second approach a given observable is interpolated using a functional
form of $O(T)=A(T)+B(T)/N_t^2$ where $A(T)$ and $B(T)$ are cubic spline functions.
The two splines have the same node points. These, however, do not coincide
with the data points. The number of node points is smaller than the number of
data points and the $A(T)$ and $B(T)$ splines are fitted to the data. The 
node points are scattered randomly in the temperature range with the following
constraints: there is always a node point before the first and after the 
last data point and in each interval between adjacent data points there can be
0 or 1 node point. In order to determine the systematic uncertainty of the
resulting $O(T)$ curve we weight the various splines using the Akaike Information
Criterion (AIC)~\cite{Akaike1973,Akaike1978c,Borsanyi:2014jba}. 
The weight of each fit result is $w_i=\exp(AIC_i/2)$ with $AIC_i=2k_i-\chi_i^2$
where $k_i$ is the number of degrees of freedom and $\chi_i^2$ is the usual $\chi^2$
for the $i^{\rm th}$ fit.
Since there are many possibilities to select node points instead of
evenly distributing them we used importance sampling based on the AIC
weights. Thus starting from a random set of node points at each step we propose
one of the following changes: adding a new node point, removing one, or shifting
one without breaking the above constraints. Then the proposed set of
new node points is either accepted or rejected using a Metropolis step with
probability: $p=\min\left\{ 1, \exp(\Delta AIC/2)\right\}$.
Since the number of degrees of freedom $k_i$ (i.e. the number
of node points) can change during the Monte-Carlo sampling of node points, bad
fit qualities result in small $k_i$. In the extreme case when the data 
is highly inconsistent with the above functional form 
(i.e. the lattice spacings
are far from the scaling regime) the $AIC$ weights are maximized by having no
degree of freedom and consequently vanishing $\chi^2$. We do not observe such
behavior for any of our observables, the required number of degrees
of freedom is always ${\cal O}(10)$, proving that the continuum 
extrapolations are under control. The resulting $O(T)$ curves can
simply be averaged and the width of their
distribution defines our systematic uncertainty. Statistical errors are determined
by a jackknife analysis. 

The two analyses gave consistent results in all cases. The 
results presented in the following were obtained with the second one.

As mentioned previously the staggered ensembles were tuned to have the same
temperatures for all $N_t$. This makes a pointwise continuum extrapolation
trivial. The systematic error has been estimated from continuum extrapolations
using all four lattice spacings or only the finest three.

\subsection{Observables}

\begin{figure}[h]
    \centering
    \includegraphics[width=12cm]{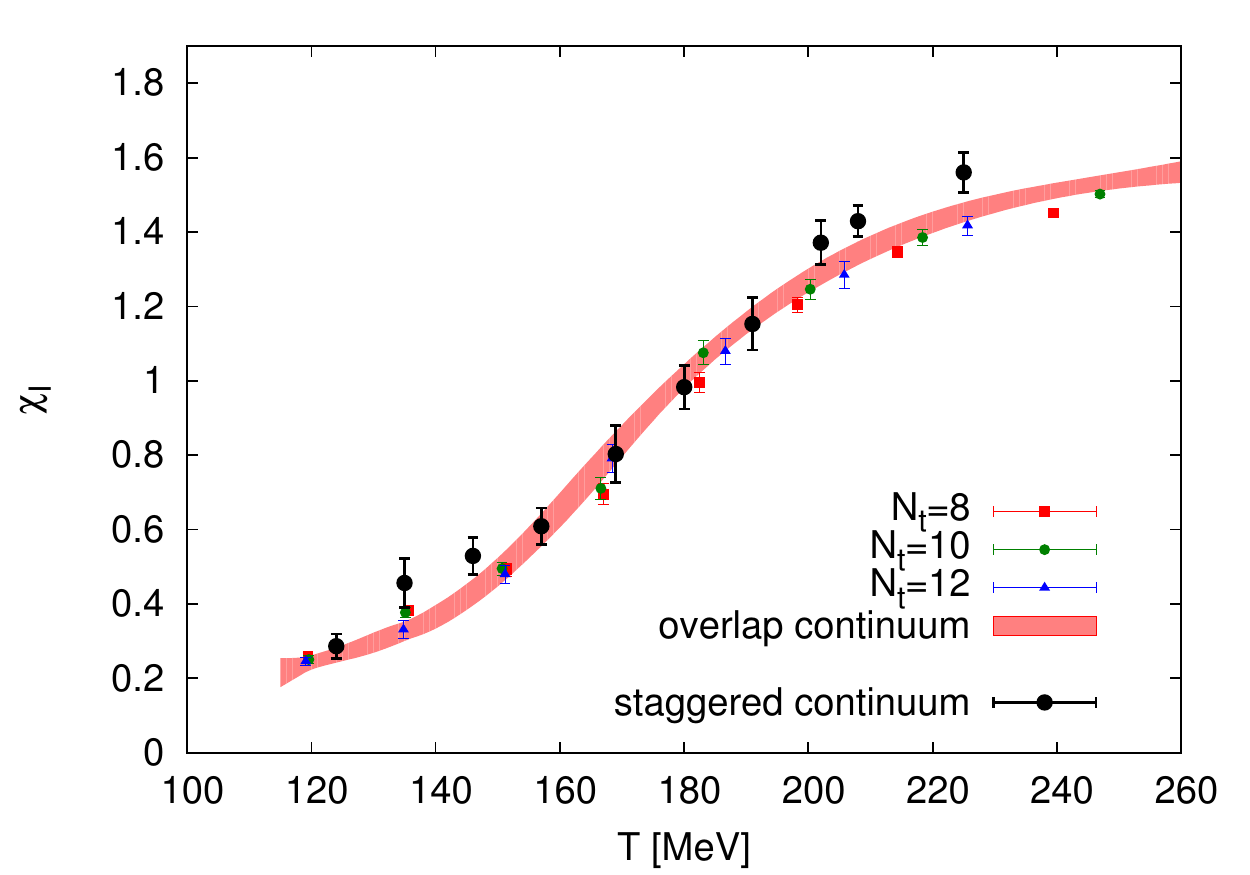}
    \caption{\label{fi:qns} Isospin susceptibility as a function of temperature.
    The red, green, and blue symbols show the overlap results on $N_t=$8, 10, 12 lattices
    and the purple band is their continuum extrapolation. The continuum extrapolated
    staggered result using only the $Q=0$ configurations is shown by the black dots.}
\end{figure}

The first observable we determine is the isospin susceptibility which is the
connected part of the quark number susceptibility, defined as:
\[ \chi_I=\frac{T}{V}\left.\frac{\partial^2}{\partial \mu_I^2}\right|_{\mu_I=0} \log Z\]
We perform a tree level improvement using the correction factors listed in Table 1
of~\cite{Borsanyi:2012xf}. Figure \ref{fi:qns} shows the results on the 
three lattice spacings and their continuum extrapolation as well as the staggered
continuum result using only the $Q=0$ configurations. 
There is a nice agreement between the two discretizations. 

\begin{figure}[h]
    \centering
    \includegraphics[width=12cm]{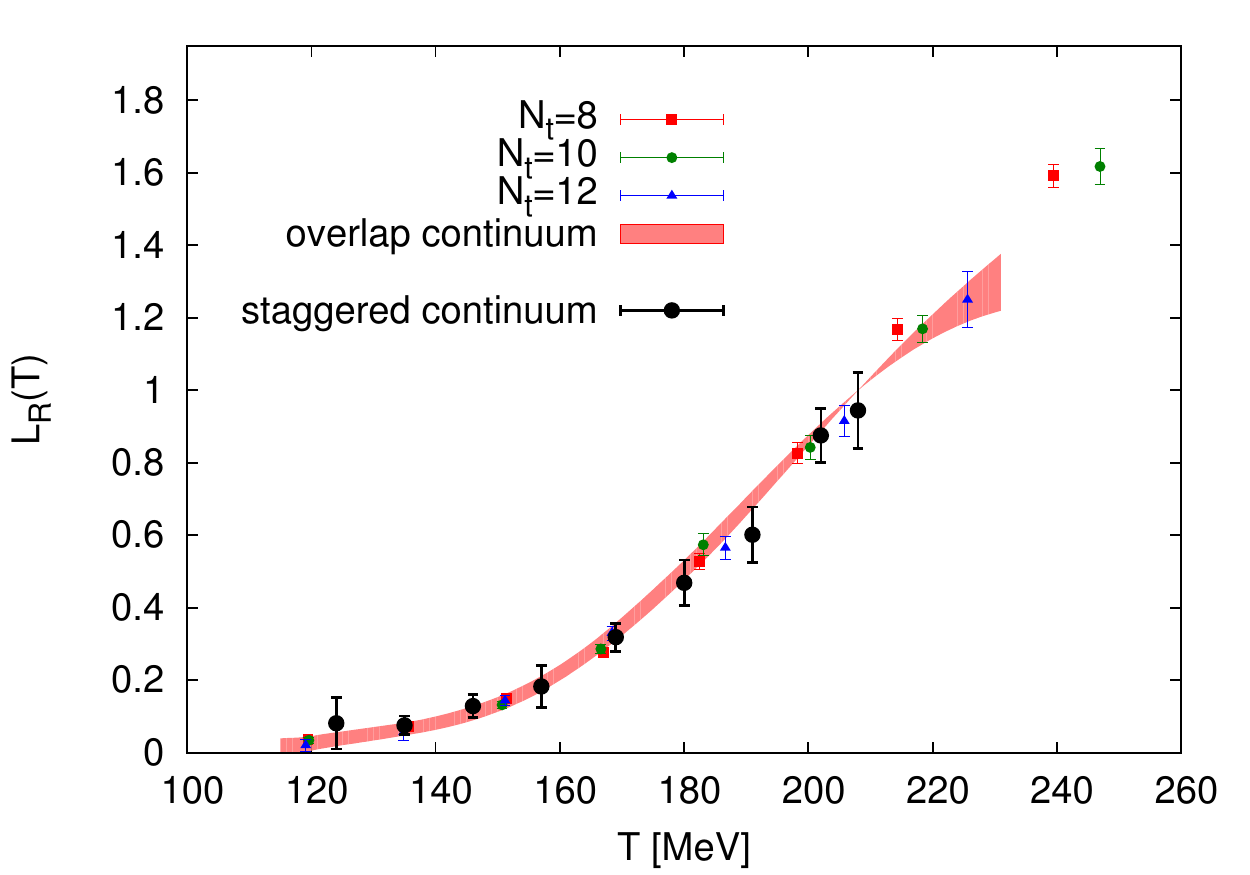}
    \caption{\label{fi:ploop} Polyakov loop as a function of
    temperature. The symbols are the same as in Figure~\ref{fi:qns}. Since no
    topology dependence is observed the staggered result contains all topological
    sectors.}
\end{figure}

The second observable is the Polyakov loop. The renormalization condition is $\left.L_R\right|_{T=208{\rm MeV}}=1$, according
to which we have
\[ L_R= L \exp\left(-N_t F_0(\beta)\right), \]
where the determination of $F_0(\beta)$ is described in Section \ref{se:ov}. It is shown on Figure \ref{fi:ploop} again
together with the staggered result. In case of the Polyakov loop the selection
of $Q=0$ configurations at low temperatures causes a significant 
loss of statistics at $N_t=16$
which results in large errors after continuum extrapolation. We checked on
the $N_t=8$ and 10 lattices that this selection has no effect on the result, 
therefore
we show the continuum extrapolated Polyakov loop obtained from all sectors.
The continuum extrapolations of the two discretizations are again consistent with each other.

\begin{figure}[h]
    \centering
    \includegraphics[width=12cm]{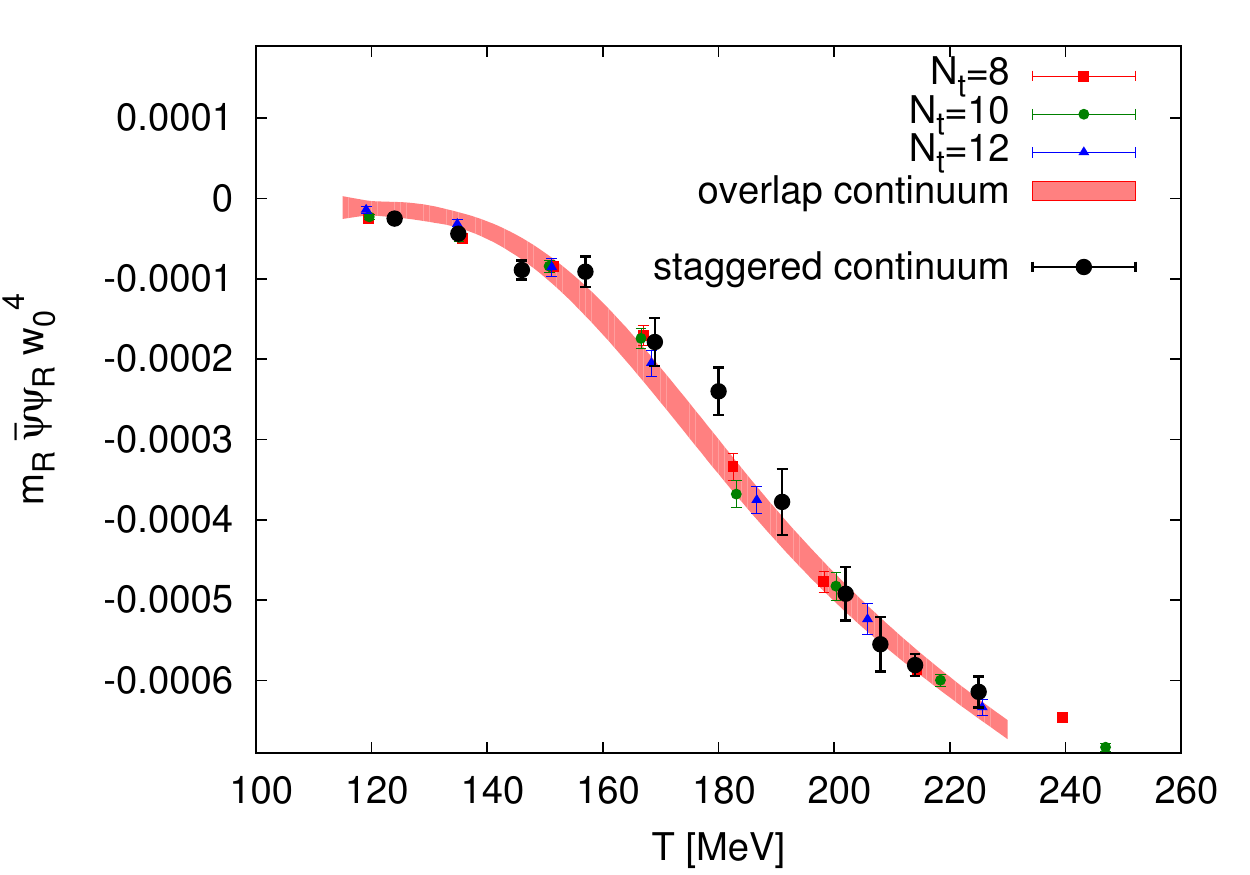}
    \caption{\label{fi:pbp} Chiral condensate as a function of
    temperature. The symbols are the same as in Figure~\ref{fi:qns}. The staggered
    result corresponds to $Q=0$.}
\end{figure}

The third observable is the chiral condensate. On Figure \ref{fi:pbp} we show
\[ m_R \overline{\psi}\psi_R w_0^4= m \left[ \frac{T}{V}\frac{\partial}{\partial m} \log Z\right]_{\rm sub}, \]
where $[\dots]_{\rm sub}$ means the zero temperature subtracted value. The staggered
result again corresponds to the $Q=0$ topological sector.

\begin{figure}[h]
    \centering
    \includegraphics[width=12cm]{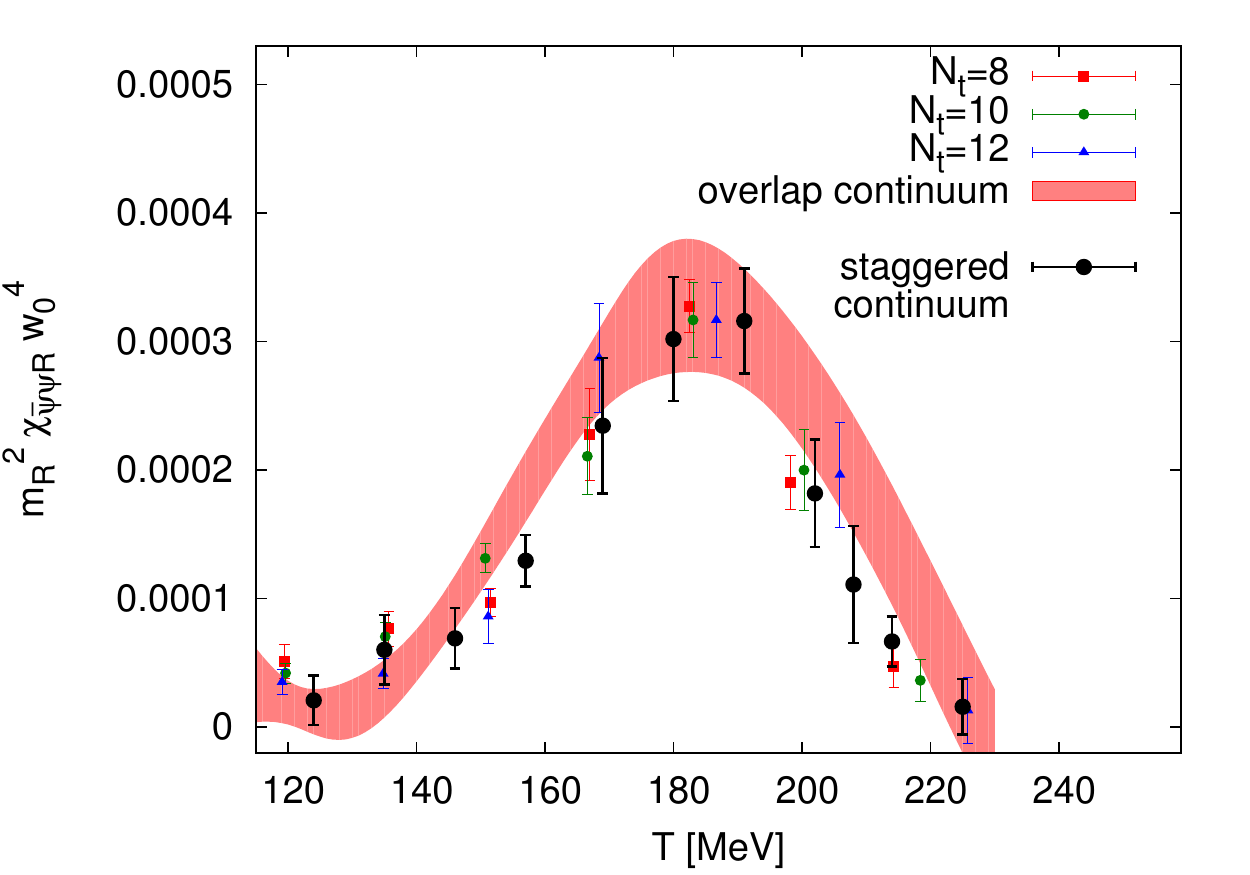}
    \caption{\label{fi:psusc} Chiral susceptibility as a function of
    temperature. The symbols are the same as in Figure~\ref{fi:qns}. The staggered
    result corresponds to $Q=0$.}
\end{figure}

The fourth observable is the chiral susceptibility. On Figure \ref{fi:psusc} we show
\[ m_R^2 \chi_{\overline{\psi}\psi R} w_0^4= m^2 \left[ \frac{T}{V}\frac{\partial^2}{\partial m^2} \log Z\right]_{\rm sub} \]
obtained both from the overlap and staggered ensembles. For both chiral observables
we see again a nice agreement between overlap and staggered results.

\begin{figure}[h]
    \centering
    \includegraphics[width=12cm]{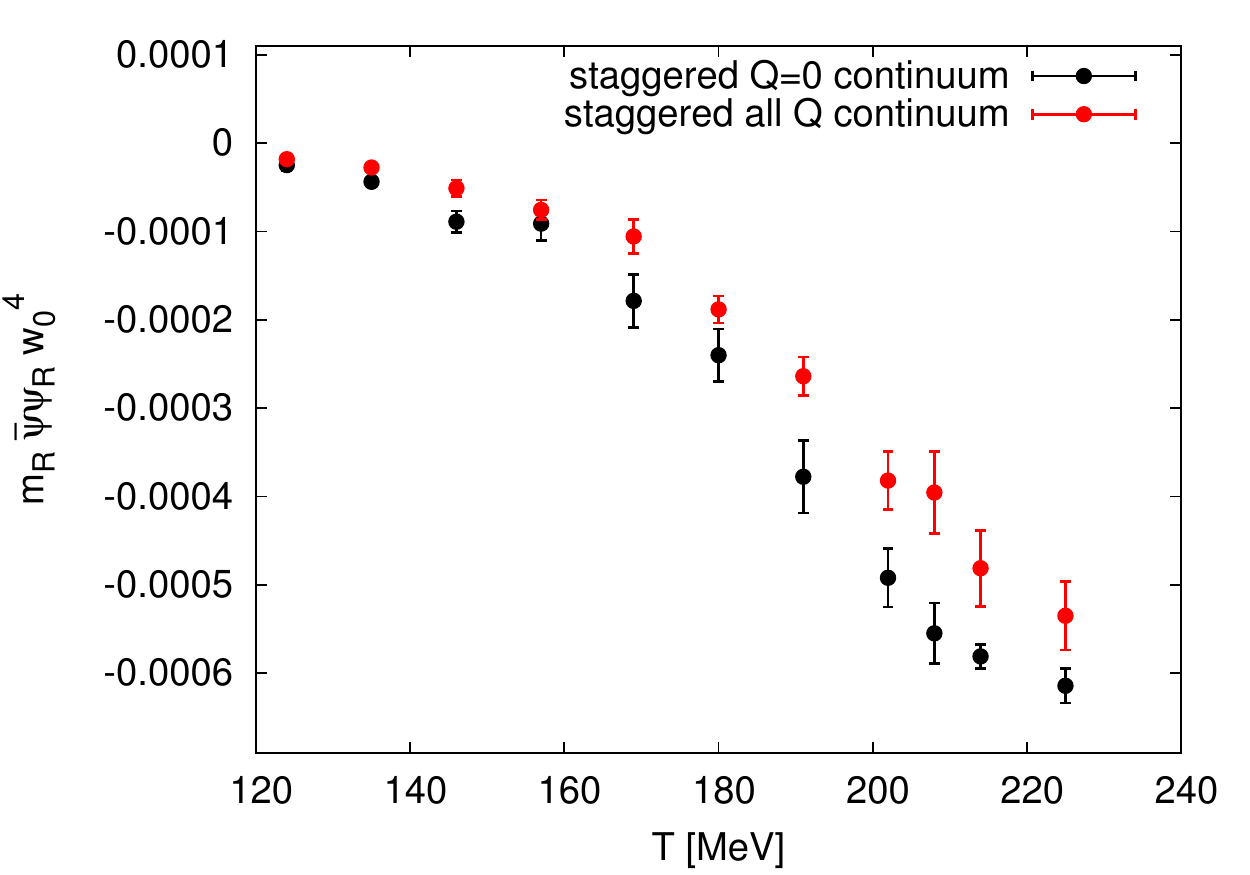}
    \caption{\label{fi:pbpq} Comparison of the staggered continuum extrapolated
    chiral condensate in the $Q=0$ sector and in all sectors.}
\end{figure}

Let us now discuss the effect of fixed topology. Using the full staggered ensembles
we can quantify how much our observables depend on the global topology in these
relatively small volumes. The quark number susceptibility and Polyakov loop are
completely insensitive, using the full ensembles give consistent results
but with smaller errors due to using the full statistics. The chiral observables, however show a different 
behavior, there is a significant difference between using $Q=0$ and using all
sectors. This is demonstrated in Figure~\ref{fi:pbpq} where the continuum
extrapolated chiral condensate is shown for these two ensembles. The difference
is expected to scale with $1/V$ therefore using an aspect ratio of $r=4$ which
is typical for previous staggered studies, it is smaller 
by an order of magnitude.

\section{Conclusions}

We have presented continuum extrapolated results for the temperature dependence
of four observables, the isospin susceptibility, the Polyakov loop, the chiral condensate,
and the chiral susceptibility using two flavors of dynamical overlap fermions
with a pion mass of $\approx$350~MeV. Three temporal extents, $N_t=$8, 10 and 12 
were used which made a controlled continuum extrapolation possible. All runs were performed at fixed topology.
The results were compared to continuum extrapolated staggered ones and a nice
agreement was found for all observables when the topology was also constrained
on the staggered ensembles. Using the full staggered ensembles as well we could 
estimate the finite volume effect caused by fixing the topology. The isospin 
susceptibility and the Polyakov loop show no $Q$ dependence already on these
relatively small volumes. The chiral observables, on the other hand, still depend
on $Q$. It is also possible to
perform simulations in different sectors and use the topological susceptibility
to combine them~\cite{Egri:2005cx}.

\section{Acknowledgments}

We thank G. Bali, T.W. Chiu, G. Endrodi and A. Schaefer for useful discussions.
Computations were performed on JUQUEEN at FZ-J\"ulich and on GPU clusters
at Wuppertal and Budapest. This project was funded by the DFG grant SFB/TR55, by OTKA under grants
OTKA-NF-104034 and OTKA-K-113034. S.D.K. is funded
by the "Lend\"ulet" program of the Hungarian Academy of Sciences (LP2012-44/2012).

\appendix
\section{Methods and algorithms}

\subsection{Preconditioning the inverse of the overlap matrix}
\label{se:prec}

Beyond the algorithmic ingredients already applied in
Ref~\cite{Borsanyi:2012xf}, i.e. Hasenbusch trick~\cite{Hasenbusch:2001ne},
Omelyan integrator~\cite{Takaishi:2005tz} and a multi-scale
scheme~\cite{Sexton:1992nu}, a major algorithmic improvement in this work is
the preconditioner for the inversion of the overlap matrix. The technique was
proposed in~\cite{Brannick:2014vda}. 

The inverter is the FGMRES algorithm. The preconditioner is the inverse of the
Wilson operator with a mass, that is a tuneable parameter. We add a clover term
to the Wilson operator, since it makes the preconditioning more efficient: it
brings the physical spectrum of the Wilson operator closer to the spectrum of
the overlap operator. We utilize an even-odd preconditioned BICGSTAB inverter
to invert the Wilson matrix. Since it is still a subdominant part of the total
inverter, we did not implement the multigrid acceleration for the Wilson
inverse, that was proposed in ~\cite{Brannick:2014vda}. The preconditioner
utilizes single precision arithmetics.

In the HMC update the inverse of the overlap operator is required at three
different occasions: in the heatbath, in the pseudofermion action and in the
pseudofermion force.  We use two flavors, so in principle the inverse of the
square of the overlap operator $(D^{\dagger}D)^{-1}$ is needed.  The cases of
the heatbath and the pseudofermion action simplify to the application of only
$D^{-1}$ for which we can use the preconditioned inverter.  In the case of the
fermion force we use two consecutive application of the preconditioned
inverter: $(D^{\dagger}D)^{-1}= D^{-1}\gamma_5D^{-1} \gamma_5$.  Even this two
step approach is about a factor five faster compared to the previously used
technique (relaxed CG).

\subsection{Algorithm to determine the index}

\newcommand{\epsa}{\varepsilon_{\text{stop}}}
\newcommand{\epsn}{\varepsilon_{\text{zero}}}
\newcommand{\epsp}{\varepsilon_{\text{nonzero}}}

Although the action is designed to keep the overlap simulations
in fixed sectors, due to the finite stepsize HMC integration sector
changes are in principle possible. Therefore it is important to monitor
that topology is indeed fixed. Below we discuss how we determined
the topological charge using the index of the overlap operator.
\algnewcommand\algorithmicto{\textbf{to}}
\algrenewtext{For}[3]%
{\algorithmicfor\ $#1 \gets #2$ \algorithmicto\ $#3$ \algorithmicdo}

\begin{algorithm}
\caption{Inverse iteration to find index of overlap operator}
\label{alg:index}
\begin{algorithmic}[1]
\Procedure{Index}{$\sigma,\epsp,\epsn,\epsa$}
\State $i\gets 0$
\Comment{The index will be stored in $i$.}
\State $n_{\text{zero}} \gets 0$, $n_{\text{vec}} \gets 0$
\Comment{Start without vectors.}
\Repeat
\If{$n_{\text{zero}} = n_{\text{vec}}$}
\Comment{If all vectors are zero modes,}
\State $n_{\text{vec}} \gets n_{\text{vec}} + 1$
\Comment{introduce a new vector.}
\State $v_{n_{\text{vec}}} \gets \text{Gaussian random vector}$
\State $\displaystyle v_{n_{\text{vec}}} \gets v_{n_{\text{vec}}} -
\sum_{l=1}^{n_{\text{vec}}-1} \langle v_l | v_{n_{\text{vec}}} \rangle\, v_l$
\State $\displaystyle v_{n_{\text{vec}}} \gets \frac{v_{n_{\text{vec}}}}{
  \left\| v_{n_{\text{vec}}} \right\|}$
\Comment{Orthogonalize w.r.t.\ previous vectors.}
\State $\varepsilon_{n_{\text{vec}}} \gets \epsp+1$
\EndIf
\For{k}{1}{n_{\text{vec}}}
\If{$\varepsilon_{k} > \epsa$}
\Comment{Don't update if vector is too precise.}
\State $\displaystyle w \gets \left( D_0^{\dagger} D_0 + \sigma \right)^{-1} v_k$
\Comment{Invert using FGMRES.}
\label{alg:index:inv}
\State $\mu \gets \langle w | v_k \rangle$
\State $\displaystyle \lambda_k \gets -\sigma + \frac{1}{\mu}$
\Comment{Estimate eigenvalue.}
\State $\displaystyle \varepsilon_k \gets \left\| v_k - \frac{w}{\mu} \right\|
/ \left\| w \right\|$
\Comment{Estimate error of eigenvalue.}
\State $\displaystyle w \gets w - \sum_{l=1}^{k-1} \langle v_l | w
\rangle \, v_l$
\State $\displaystyle v_k \gets \frac{w}{ \left\| w \right\|}$
\Comment{Keep orthonormality of vector set.}
\EndIf
\EndFor
\If{$\left( \varepsilon_{n_{\text{vec}}} \le \epsn \right) \wedge \left(
  \lambda_{n_{\text{vec}}} < \varepsilon_{n_{\text{vec}}} \right)$}
\State $n_{\text{zero}} \gets n_{\text{zero}} + 1$
\Comment{If a zero mode is encountered,}
\State $\displaystyle i \gets i - \mathrm{sgn}\big[ \langle
  v_{n_{\text{zero}}} | \gamma_5 | v_{n_{\text{zero}}} \rangle \big]$
\Comment{update the index.}
\EndIf
\Until{$\left(\varepsilon_{n_{\text{vec}}} \le \epsp \right) \wedge \left(
  \lambda_{n_{\text{vec}}} > \varepsilon_{n_{\text{vec}}} \right)$}
\Comment{Found a nonzero eigenvalue.}
\State \Return $i$
\EndProcedure
\end{algorithmic}
\end{algorithm}

The topological charge of a gauge configuration is given by the index $Q= n_- -
n_+$ of the massless overlap operator $D_0$, where $n_-$ and $n_+$ denote the
number of zero modes with negative and positive chirality, respectively.  We
systematically look for the lowest eigenvalues of $D_0^\dagger D_0$, and stop when
the first nonzero eigenvalue is found. 

The method we apply to find the zero modes, described in Algorithm
\ref{alg:index}, is based on the inverse iteration. For this one inverts the
$D_0^\dagger D_0$ with a shift $\sigma$, which is a fixed parameter of the algorithm.
There are three parameters controlling the precision of the eigenmodes, they
have to obey the constraint $0 < \epsa \le \epsn \le \epsp$. Parameter $\epsa$
controls the condition when an eigenvector is considered precise enough for the
algorithm to stop iterating over it. It is still used for the orthogonalization
process. A vector is considered as a zero mode if the corresponding eigenvalue
is compatible with zero with an error less than $\epsn$. A vector is considered
as nonzero if the corresponding eigenvalue is different from zero with an error
less than $\epsp$.

The inversion in line \ref{alg:index:inv} of Algorithm \ref{alg:index} is
computed as two consecutive inversions:
\begin{equation}
    \left( D_0^\dagger D_0 + \sigma \right)^{-1} = \left( D_0 + i\gamma_5\sqrt{\sigma}
    \right)^{-1} \gamma_5 \left( D_0 - i\gamma_5\sqrt{\sigma} \right)^{-1} \gamma_5.
\end{equation}
The inversions of $D_0 \pm i\gamma_5\sqrt{\sigma}$ are
performed using the FGMRES inverter, as described in Subsection \ref{se:prec}.
For the shift $\sigma$ we chose
a positive value, tuned such that it allows for a rapid convergence in the
inversions, while keeps the number of iterations in Algorithm \ref{alg:index}
as low as possible. In addition, a positive value of $\sigma$ ensures that the
zero eigenmodes are found first.

As an example, using the parameter 
values $\epsa=10^{-8}$, $\epsn=10^{-6}$, $\epsp=10^{-4}$ and 
$\sigma=10^{-4}$ on a $12 \times 24^3$
lattice with $Q=-3$, the algorithm 
found the three zero modes and the first nonzero mode within 8 iterations.

\section*{References}

\bibliographystyle{elsarticle-num}
\bibliography{ov15}

\end{document}